\documentclass[aps,prl,showpacs]{revtex4}
\usepackage{amssymb}
\usepackage{amsmath}

\usepackage{graphicx}
\usepackage{bm}

\begin{document}

\title{Explicit Breaking of $SO(3)$ with Higgs Fields in the Representations 
$\ell =2$ and $\ell =3$}
\date{\today}
\author{Mehmet Koca}
\email{kocam@squ.edu.om}
\affiliation{Department of Physics, College of Science, Sultan Qaboos University, PO Box
36, Al-Khod 123, Muscat, Sultanate of Oman}
\author{Ramazan Ko\c{c}}
\email{koc@gantep.edu.tr}
\affiliation{Department of Physics, Faculty of Engineering University of Gaziantep, 27310
Gaziantep, Turkey}
\author{Hayriye T\"{u}t\"{u}nc\"{u}ler}
\email{tutunculer@gantep.edu.tr}
\affiliation{Department of Physics, Faculty of Engineering University of Gaziantep, 27310
Gaziantep, Turkey}

\begin{abstract}
A gauged SO(3) symmetry is broken into its little groups of the
representations $\ell $=2 and $\ell $=3. Explicit Higgs potentials leading
to the spontaneous symmetry breaking are constructed. The masses of the
gauge bosons and Higgs particles are calculated in terms of the
renormalizable potentials. Emergence of Goldstone bosons arising from the
absence of certain potential terms is also discussed. Analogous structures
between the cosmic strings and disclinations of liquid crystals are noted.
\end{abstract}


\maketitle

\section{Introduction}

A general classification of little groups of $SO(3)$ has been given for its
irreducible representations\cite{1}. For $\ell =2$ they are the dihedral
groups $D_{\infty }$ and $D_{2}$, and for $\ell =3$ representations they are 
$C_{\infty }\approx SO(2)$, tetrahedral group $T$, dihedral group $D_{3}$,
the cyclic groups $C_{3}$ and $C_{2}$. They have been obtained purely from
the algebraic arguments which states that a little group $H$ should have
sufficient number of trivial representations in a given irreducible
representation of the parent group $G$. These arguments should be justified
by writing explicit potentials and the nature of the parameters associated
with the quadratic, third order and quartic terms should be clarified. As we
will see not all these closed subgroups of $SO(3)$ are realized as little
groups. The potential leading to the little group solutions \ $C_{3}$ and $%
C_{2}$\ are identical for the solutions of the tetrahedral group $T$ and the
dihedral group $D_{3}$ respectively.

In this paper we address ourselves to this problem which is rather important
in the liquid crystal phase transitions \cite{2}. There of course, the
potential term without gauge interaction plays the principal role where a
symmetric tensor field is invoked as an order parameter. In spite of the
close resemblance between two problems our main interest rests on the gauge
nature of our problem and the role of the Higgs fields in such a theory. In
an earlier paper \cite{3}, given the representation of $SO(3),$ we have
discussed how one can obtain the explicit matrix generators of the little
groups. Identification of the Higgs fields receiving non-zero vacuum
expectation values for particular little groups was also discussed. In what
follows we write down explicit Higgs potentials for each representation $%
\ell =2$ and $\ell =3$ and discuss the minimization conditions which also
lead to the masses and associate the fields with real Higgs scalars. We also
illustrate the relation between the real scalar Higgs fields and the
symmetric tensor fields which seem to be more useful in liquid crystal
phenomena \cite{4}.\newline
In chapter 2 we briefly discuss the case of $\ell =2$, a phenomena which is
widely known in physics literature \cite{5}. Chapter 3 involves the details
of symmetry breaking of $SO(3)$ with the Higgs fields in $\ell =3$
representation. Finally we discuss our results in Chapter 4 and remark on
the possible use of our method for model builders regarding particle physics
and cosmology.

\section{General framework and the Higgs scalars in $\ell $=2 representations%
}

The standard Lagrangian of a local gauge theory without fermions is given by%
\begin{equation}
L=-\frac{1}{4}F_{\mu \nu }F^{\mu \nu }-\frac{1}{2}(D_{\mu }\phi )^{+}(D_{\mu
}\phi )-V(\phi )
\end{equation}%
where the field strengths $F_{\mu \nu }$ and the covariant derivative $%
D_{\mu }$ are given by%
\begin{eqnarray}
F_{\mu \nu } &=&\partial _{\mu }W_{\nu }-\partial _{\nu }W_{\mu }+gW_{\mu
}\times W_{\nu }  \notag \\
D_{\mu } &=&\partial _{\mu }-igW_{\mu },\quad W_{\mu }=\vec{J}\cdot 
\overrightarrow{W}_{\mu }
\end{eqnarray}%
where $SO(3)$ generators $\vec{J}$ are the $(2\ell +1)\times (2\ell +1)$
matrices for the irreducible representation $\ell $. The Higgs scalars $\phi
(\ell m)$ transform like spherical harmonics $Y_{\ell m}$ under the group
transformations. We will rather prefer the real scalars $\chi _{i}$ $%
(i=1,2,\dots ,2\ell +1)$ which can be defined from the complex fields $\phi
(\ell m)$, that we will illustrate explicitly when they are needed. A
general Higgs potential restricted by renormalizability can be written as%
\cite{6}%
\begin{equation}
V(\chi )=a\chi _{i}\chi _{i}+bf_{ijk}\chi _{i}\chi _{j}\chi
_{k}+cg_{ijkl}\chi _{i}\chi _{j}\chi _{k}\chi _{l}.
\end{equation}%
Explicit forms of numerical tensors $f_{ijk}$ and $g_{ijkl}$ depend on the
representations. For $\ell =2$ representation (3) takes the form%
\begin{align}
V(\chi )& ={}a(\chi _{1}^{2}+\cdots +\chi _{5}^{2})  \notag \\
& +b[\sqrt{3}\chi _{1}(\chi _{3}^{2}-\chi _{4}^{2})+2\sqrt{3}\chi _{2}\chi
_{3}\chi _{4}  \notag \\
& -(2\chi _{1}^{2}+2\chi _{2}^{2}-\chi _{3}^{2}-\chi _{4}^{2}-\frac{2}{3}%
\chi _{5}^{2})\chi _{5}]  \notag \\
& +c(\chi _{1}^{2}+\cdots +\chi _{5}^{2})^{2}
\end{align}%
where the real fields $\chi $$_{i}$ $(i=1,\dots ,5)$are related to the
Complex Higgs scalars by the relations%
\begin{align}
\chi _{1}& =\frac{1}{\sqrt{2}}(\phi (22)+\phi (2-2)),\quad \chi _{3}=\frac{1%
}{\sqrt{2}}(\phi (21)-\phi (2-1))  \notag \\
\chi _{2}& =\frac{i}{\sqrt{2}}(\phi (22)-\phi (2-2)),\quad \chi _{4}=\frac{i%
}{\sqrt{2}}(\phi (21)+\phi (2-1))  \notag \\
\chi _{5}& =\phi (20).
\end{align}%
The potential (4) in terms of the symmetric tensor field $T_{ab}$ $%
(a,b=1,2,3),$ $T_{aa}=0$ (sum over a is understood) is used to describe the
fields in $\ell =2$ representation. For a further consideration we give the
relations between the$\chi $-fields and the components of $T_{ab}$:%
\begin{eqnarray}
T_{11} &=&\frac{1}{\sqrt{2}}\chi _{1}-\frac{1}{\sqrt{6}}\chi _{5},\quad
T_{12}=\frac{1}{\sqrt{2}}\chi _{2},  \notag \\
T_{22} &=&-\frac{1}{\sqrt{2}}\chi _{1}-\frac{1}{\sqrt{6}}\chi _{5},\,T_{23}=%
\frac{1}{\sqrt{2}}\chi _{4},  \notag \\
\,T_{33} &=&\frac{2}{\sqrt{6}}\chi _{5},\quad T_{13}=\frac{1}{\sqrt{2}}\chi
_{3},
\end{eqnarray}%
The potential (4) in terms of the field $T_{ab}$ would read%
\begin{equation}
V(T)=aTrT^{2}+\sqrt{\frac{8}{3}}bTrT^{3}+c\left[ TrT^{2}\right] ^{2}.
\end{equation}%
The advantage of (4) is that the potential is expressed in terms of
independent fields $\chi $$_{i}$ whereas (7) includes also dependent fields.
Moreover, (4) is more convenient for a gauge theory.\newline
\qquad Let us consider the little groups $D_{2}$ and $D_{\infty }$ of $\ell
=2$ in turn. Before we proceed further we discuss the general character of
symmetry breaking mechanism for a general potential. The spontaneous
symmetry-breaking takes place if the minimality conditions 
\begin{subequations}
\begin{eqnarray}
\left\langle \frac{\partial V}{\partial \chi _{i}}\right\rangle &=&0 \\
\left\langle \frac{\partial ^{2}V}{\partial \chi _{i}\partial \chi _{j}}%
\right\rangle &\geq &0\quad (i,j=1,2\ldots ,2\ell +1)
\end{eqnarray}%
are satisfied for a set of physical parameters a, b, and c. (8.a-b)
determine not only the range of parameter a, b, and c should hold but the
eigenvalues of the mass-squared matrix 
\end{subequations}
\begin{equation}
M_{ij}^{2}=\frac{1}{2}\left\langle \frac{\partial ^{2}V}{\partial \chi
_{i}\partial \chi _{j}}\right\rangle
\end{equation}%
which yield to the masses of the Higgs fields.

\subsection{The Little group D$_{2}$}

The $\ell =2$ representation has two trivial representations of $D_{2}$
which can be associated with the fields \textmd{\textup{$\chi _{1}$}} 
\textmd{\textup{and}} \textmd{\textup{$\chi _{5}$}}\cite{3}\textmd{\textup{.
When these fields take non-zero vacuum expectation values}}%
\begin{equation}
\left\langle \chi _{1}\right\rangle =v_{1}\quad and\quad \left\langle \chi
_{5}\right\rangle =v_{5}
\end{equation}%
it is expected that $SO(3)$ breaks into $D_{2}$. (8a) leads to two sets of
independent equations%
\begin{eqnarray*}
av_{1}-2bv_{1}v_{5}+2cv_{1}(v_{1}^{2}+v_{5}^{2}) &=&0 \\
av_{5}+b(v_{5}^{2}-v_{1}^{2})+2cv_{5}(v_{1}^{2}+v_{5}^{2}) &=&0
\end{eqnarray*}%
Assuming that both $v_{1}\neq 0$ and $v_{5}\neq 0$ we obtain%
\begin{equation}
bv_{1}(3v_{5}^{2}-v_{1}^{2})=0.
\end{equation}%
If $v_{1}\neq 0$ and/or $3v_{5}^{2}\neq v_{1}^{2}\neq 0$ are invoked then $%
b=0$ should necessarily hold. This choice of vacuum expectation values
transform $T_{ab}$ to a diagonal matrix which is equivalent to a $SO(3)$
gauge fixing. As for the eigenvalues of real traceless symmetric matrix they
come in two classes, degenerate and non-degenerate cases, which put certain
restrictions on the vacuum expectation values $v_{1}$ and $v_{5}$. The
non-degenerate eigenvalues of $T_{ab}$ can be written as $(1,-1,0)v_{1}/%
\sqrt{2}$ which is equivalent to taking $v_{1}\neq 0$ and $v_{5}=0$. Of
course, alternative choices $(0,1,-1)v_{1}/\sqrt{2}$ and $(-1,0,1)v_{1}/%
\sqrt{2}$ which restrict $v_{5}$ to take $v_{5}=\pm \sqrt{3}v_{1}.$ Either
of this choice require $b=0$. Therefore we can work out without loss of
generality with the case of $v_{1}\neq 0$ and $v_{5}=0$ where the $\chi $$%
_{1}$ field is trivial representation of $D_{2}$ but not of $D_{\infty }$.
This indicates that in breaking $SO(3)$ into $D_{2}$ the third order term in
potential $V(\chi )$ should be absent. (10a) leads to the condition $%
v_{1}^{2}=-\frac{a}{2c}>0.$ Then the masses of Higgs fields are%
\begin{equation}
m_{1}^{2}=m_{2}^{2}=m_{3}^{2}=m_{4}^{2}=0,\quad \quad m_{5}^{2}=-2a>0
\end{equation}%
which, with $-\frac{a}{2c}>0,$ implies that in the potential (4) we should
have $a<0,b=0$ and $c>0$. This is the unique solution for $SO(3)\rightarrow
D_{2}$ breaking. We note that the gauge bosons gain the masses%
\begin{equation}
M_{W^{\pm }}=g\sqrt{\frac{-a}{2c}},\quad M_{W^{0}}=2M_{W^{\pm }}.
\end{equation}%
Three of the Higgs fields are absorbed by the gauge bosons. Then it is
natural to obtain three zero eigenvalues from the mass matrix (9). However
we obtain four zero eigenvalues which indicates that one of the Higgs
scalars remain as a Goldstone boson while the others gain a mass of $\sqrt{%
-2a}.$

\subsection{The little group D$_{\infty }$ of $\ell $=2}

Here only $\chi $$_{5}$ transforms as a trivial representation of $D_{\infty
}$ \cite{3}. When we take the vacuum expectation values $\left\langle \chi
_{i}\right\rangle =0,\,\,i\neq 5\;$and$\quad \left\langle \chi
_{5}\right\rangle =v_{5}\neq 0$ we obtain from (8a)

\begin{equation}
a+bv_{5}+2cv_{5}^{2}=0.
\end{equation}%
The case $v_{1}=0\quad $and$\quad v_{5}\neq 0$ certainly corresponds to the
degenerate eigenvalues of $T:\,(-1,-1,2)\frac{v_{5}}{\sqrt{6}}.$ The
alternative solutions $3v_{5}^{2}=v_{1}^{2}$ consistent with $b\neq 0$ are
just reshuffling the orders of the eigenvalues as $(2,-1,-1)\frac{v_{5}}{%
\sqrt{6}}\quad and\quad (-1,2,-1)\frac{v_{5}}{\sqrt{6}}.$ They correspond to
a definition of a D$_{\infty }$ trivial representation as a linear
combination of $\chi $$_{1}$ and $\chi $$_{5}$ fields. Therefore it is quite
appropriate to work with $v_{5}\neq 0\quad $and$\quad v_{1}=0.$ Now the
eigenvalues of (9) are obtained as follows%
\begin{eqnarray}
m_{1}^{2} &=&m_{2}^{2}=0  \notag \\
m_{3}^{2} &=&m_{4}^{2}=-3bv_{5}>0 \\
m_{5}^{2} &=&v_{5}(b+4cv_{5})>0  \notag
\end{eqnarray}%
It is clear from (15) that two of the Higgs scalars are absorbed by gauge
fields giving them the masses 
\begin{equation}
M_{W^{\pm }}=g\sqrt{3}v_{5},\quad M_{W^{0}}=0.
\end{equation}%
For $ac\mathtt{>}0$, from (14,15,16) we deduce that

\begin{equation}
c\mathtt{>}0,b\mathtt{<}0,a\mathtt{>}0,b^{2}\geq 8ac\text{ \ and \ \ }v_{5}=-%
\frac{b}{4c}+\frac{\sqrt{b^{2}-8ac}}{4c}>0
\end{equation}%
or for $ac<0$ we simply have $c>0,a<0,b<0.$ We note that there is no
Goldstone boson in $SO(3)\rightarrow D_{\infty }$ breaking provided the
third order term $b\neq 0$ is included. Otherwise one obtains two Goldstone
bosons in addition to a Higgs field with mass $m_{5}^{2}=-2a.$

As we have noted the potential with $b=0$ leads to the Goldstone boson
solutions in both symmetry breaking mechanisms. This is due to the fact that
the potential with $b=0$ has a global $SO(5)$ symmetry larger than $SO(3)$
gauge symmetry\cite{col}.

\section{ Breaking SO(3) by the Higgs Scalars of $\ell $=3}

A completely symmetric tensor $T_{abc}=0(a,b,c=1,2,3)$ of rank 3 with the
trace condition $T_{aab}=0$ can be used to describe the Higgs scalars.
Although the symmetric tensor has more practical use in the liquid crystal
phenomena \cite{4} it is not very convenient for our calculations. We rather
prefer working with the $\chi _{i}(i=1,2,\dots ,7)$ fields for $\ell =3$
representation. The $\chi _{i}$ scalars can be defined from the Higgs fields 
$\phi (\ell m)$:%
\begin{eqnarray}
\chi _{1} &=&\frac{1}{\sqrt{2}}(\phi (33)-\phi (3-3)),\quad \chi _{4}=\frac{i%
}{\sqrt{2}}(\phi (32)-\phi (3-2))  \notag \\
\chi _{2} &=&\frac{i}{\sqrt{2}}(\phi (33)+\phi (3-3)),\quad \chi _{5}=\frac{1%
}{\sqrt{2}}(\phi (31)-\phi (3-1)) \\
\chi _{3} &=&\frac{1}{\sqrt{2}}(\phi (32)+\phi (3-2)),\quad \chi _{6}=\frac{i%
}{\sqrt{2}}(\phi (31)+\phi (3-1))  \notag \\
\quad \chi _{7} &=&\phi (30).  \notag
\end{eqnarray}%
It is not difficult to express the components of the symmetric traceless
tensor in terms of the $\chi $-fields%
\begin{eqnarray}
T_{112} &=&-\frac{1}{2}\chi _{2}+\frac{1}{2\sqrt{15}}\chi _{6},\quad T_{223}=%
\frac{1}{\sqrt{6}}\chi _{3}+\frac{1}{\sqrt{10}}\chi _{7},  \notag \\
T_{113} &=&-\frac{1}{\sqrt{6}}\chi _{3}+\frac{1}{\sqrt{10}}\chi _{7},\quad
\quad T_{233}=-\frac{2}{\sqrt{15}}\chi _{6},  \notag \\
T_{122} &=&\frac{1}{2}(\chi _{1}+\frac{1}{\sqrt{15}}\chi _{5}),\quad \quad
\quad \,T_{123}=-\frac{1}{\sqrt{6}}\chi _{4},\quad \quad \\
\quad T_{133} &=&-\frac{2}{\sqrt{15}}\chi _{5}.\quad  \notag
\end{eqnarray}%
The other three components of tensor field $T_{111}$, $T_{222}$, and $%
T_{333} $ can be obtained from the vanishing of trace of the tensor field.
It is easier to write down the potential with the tensor field which reads%
\begin{eqnarray}
V(T) &=&aT_{ijk}T_{ijk}+b(T_{ijk}T_{ijk})^{2}  \notag \\
&&-6c[19T_{ijk}T_{ij\ell }T_{mnk}T_{mn\ell }+44T_{ijk}T_{i\ell m}T_{j\ell
n}T_{kmn}].
\end{eqnarray}%
Note that no third order invariant polynomial exists for $\ell $=3 potential
but we have two independent fourth order polynomials. In terms of the $\chi $%
-fields the potential (20) reads explicitly%
\begin{align}
V(\chi )& ={}a(\chi _{1}^{2}+\cdots +\chi _{7}^{2})+b(\chi _{1}^{2}+\cdots
+\chi _{7}^{2})^{2}+c[9(\chi _{1}^{2}+\chi _{2}^{2})^{2}  \notag \\
& -16(\chi _{3}^{2}+\chi _{4}^{2})^{2}+18(\chi _{1}^{2}+\chi _{2}^{2})(\chi
_{3}^{2}+\chi _{4}^{2})  \notag \\
& -42((\chi _{1}^{2}+\chi _{2}^{2})(\chi _{5}^{2}+\chi _{6}^{2})-2(\chi
_{3}^{2}+\chi _{4}^{2})(\chi _{5}^{2}+\chi _{6}^{2})  \notag \\
& -72((\chi _{1}^{2}+\chi _{2}^{2})\chi _{7}^{2}+48(\chi _{3}^{2}+\chi
_{4}^{2})\chi _{7}^{2}-16\sqrt{15}\chi _{4}\chi _{5}\chi _{6}\chi _{7} 
\notag \\
& +5(\chi _{5}^{2}+\chi _{6}^{2})^{2}+20\sqrt{15}(\chi _{1}\chi _{3}^{2}\chi
_{5}-\chi _{2}\chi _{4}^{2}\chi _{6}-\chi _{1}\chi _{4}^{2}\chi _{5}+\chi
_{2}\chi _{3}^{2}\chi _{6}) \\
& +24\sqrt{15}(\chi _{1}\chi _{5}\chi _{6}^{2}-\chi _{2}\chi _{5}^{2}\chi
_{6})+8\sqrt{15}(\chi _{3}\chi _{6}^{2}\chi _{7}-\chi _{3}\chi _{5}^{2}\chi
_{7})  \notag \\
& +8\sqrt{15}(\chi _{1}\chi _{5}^{3}+\chi _{2}\chi _{6}^{3})+40\sqrt{15}%
(\chi _{1}\chi _{3}\chi _{4}\chi _{6}+\chi _{2}\chi _{3}\chi _{4}\chi _{5}) 
\notag \\
& +120(\chi _{1}\chi _{3}\chi _{5}\chi _{7}+\chi _{2}\chi _{3}\chi _{6}\chi
_{7}+\chi _{2}\chi _{4}\chi _{5}\chi _{7}-\chi _{1}\chi _{4}\chi _{6}\chi
_{7})].  \notag
\end{align}

We now, in turn, discuss the possible little group candidates $SO(2)\approx
C_{\infty },D_{3},T,C_{3}$ and $C_{2}$ of $\ell =3$ representation of $SO(3)$%
.

\subsection{The little group SO(2) of $\ell $=3\newline
}

In ref.\cite{3} We had chosen a representation where $\chi $$_{_{7}}$
transforms as a trivial representation of SO(2). To break SO(3) into SO(2)
all the fields $\left\langle \chi _{a}\right\rangle =0\;(a\neq 7)$ take zero
expectation values except $\left\langle \chi _{7}\right\rangle =v_{7}\neq 0.$%
The minimality of the potential (8) leads to the results

\begin{equation}
v_{7}^{2}=-\frac{a}{2b}>0
\end{equation}%
\begin{eqnarray}
m_{1}^{2} &=&m_{2}^{2}=\frac{36ac}{b},\quad m_{3}^{2}=m_{4}^{2}=-\frac{24ac}{%
b}>0  \notag \\
m_{5}^{2} &=&m_{6}^{2}=0,\quad \quad \quad m_{7}^{2}=-2a>0
\end{eqnarray}%
where $m_{a}^{2}\;(a=1,\cdots ,7)$ are the eigenvalues of the mass matrix
(9). It is obvious from (22-23) that no parameters exist satisfying the
positivity of masses. Therefore $SO(2)$ is not a little group of $\ell =3$
representation when $a\neq 0,\;b\neq 0\quad $and$\quad c\neq 0.$ For $a<0$, $%
b>0$ and $c=0$ $SO(3)\rightarrow SO(2)$ breaking is possible. However we
obtain four Goldstone bosons in this latter case. We also note here that the
potential with $c=0$ has a global symmetry $SO(7)$ which leads to the
Goldstone boson solutions. This is a general case for the potential with $%
c=0 $ which will repeat in the following sections.

\subsection{The little group D$_{\mathbf{3}}$\newline
}

For this little group \cite{3} it is the $\chi _{1}$ field which transforms
as a trivial representation of $D_{3}$. Assigning the vacuum expectation
values $\left\langle \chi _{1}\right\rangle =v_{1}$ we check the minimality
conditions in (8) and obtain masses of Higgs fields using (9)

\begin{equation}
v_{1}^{2}=-\frac{a}{2(b+9c)}>0
\end{equation}%
\begin{eqnarray}
m_{1}^{2} &=&-2a>0,\quad m_{2}^{2}=m_{3}^{2}=m_{4}^{2}=0  \notag \\
m_{5}^{2} &=&m_{6}^{2}=-\frac{30ac}{b+9c}>0,\;\;m_{7}^{2}=-\frac{45ac}{b+9c}%
>0.\quad
\end{eqnarray}%
It is certain that the Higgs fields $\chi _{2}$, $\chi _{3}$ and $\chi _{4}$
are absorbed by the gauge bosons giving them the masses

\begin{equation}
M_{W^{\pm }}=g\sqrt{3}v_{1},\quad M_{W^{0}}=\sqrt{6}M_{W^{\pm }}.
\end{equation}%
From (26) and (25) we obtain the relations

\begin{equation}
a<0,\quad b+9c>0\quad and\;c<0.
\end{equation}%
The relations in (26) indicate that two Higgs fields $\chi _{1}$ and $\chi
_{7}$ transforming as singlets under $D_{3}$ gain different masses while $%
\chi _{5}$ and $\chi _{6}$ transforming as a doublet of $D_{3}$ get the same
mass as expected. The minimum of the potential takes the value $V_{\min }=-%
\frac{a^{2}}{4(b+9c)}.$

Two different cases need to be discussed in this breaking:

\textbf{\textit{i) }}$a\neq 0,\;b\neq 0,\;c=0$

This shows that the symmetry breaking is possible but one will be left with
one Higgs field and three Goldstone bosons.

\textbf{\textit{ii) }}$a\neq 0,\;b=0,\;c\neq 0$

The potential does not posses minimum in this case. Consequently no symmetry
breaking takes place at all. Therefore parameters $a,b,$ and $c$ should
satisfy the conditions (27) which lead to physical theory in $%
SO(3)\rightarrow D_{3}$ breaking.

\subsection{The Tetrahedral group T as a little group of $\ell $=3
representation}

The $\ell =3$ representation of $SO(3)$ is the lowest dimensional
representation which admits the tetrahedral group $T$ as a little group of $%
SO(3)$. In the representation of ref.\cite{3} $\chi $$_{_{4}}$ is the only
scalar field which transforms as a trivial representation of the tetrahedral
group $T$. When it gains non-zero expectation value $\left\langle \chi
_{a}\right\rangle =0\;(a\neq 4)$ the $SO(3)$ is expected to break into the
subgroup $T$.

The minimality conditions (8) and the eigenvalues of Higgs masses lead to
the results

\begin{equation}
v_{4}^{2}=-\frac{a}{2(b-16c)}>0
\end{equation}%
\begin{eqnarray}
m_{1}^{2} &=&m_{2}^{2}=m_{3}^{2}=0,  \notag \\
m_{4}^{2} &=&m_{5}^{2}=m_{6}^{2}=-\frac{40ac}{b-16c}>0,\;\;m_{7}^{2}=-2a>0.
\end{eqnarray}%
Here what we observe that the squared-mass matrix is not diagonal and those
in (30) are just the eigenvalues of the matrix where the associated fields
are the linear combinations of $\left\langle \chi _{a}\right\rangle $
fields. They were indeed listed in ref. \cite{3} and we don't see any reason
to reproduce them here. The (29) and (30) are satisfied provided we have

\begin{equation}
a<0,\quad b-16c>0\quad and\;c>0.
\end{equation}%
Three of the Higgs fields are absorbed by the gauge bosons which gain equal
masses

\begin{equation}
M_{W^{\pm }}=M_{W^{0}}=\sqrt{2}gv_{4}.
\end{equation}%
The remaining three Higgs fields which transform as a three-dimensional
representation of T gain equal masses $\sqrt{\frac{40ac}{-b+16c}},$ while
the last field which transforms as a non-trivial singlet gain the mass $%
\sqrt{-2a}.$ It is quite interesting to note that (31) is different from
(28) in the sense that $c>0$ in (31) but $c<0$ in the latter. That is, in
the SO(3)$\rightarrow $D$_{3}$ breaking c is strictly negative whereas in $%
SO(3)\rightarrow T$, $c$ is strictly positive. Positive quartic terms are to
be present in the potential in order to obtain a $SO(3)$ breaking into the
tetrahedral group T.\newline
If one ignores the last fourth order term ($c=0$) in the potential which
possesses a global $SO(7)$ symmetry, the symmetry breaking is possible but
with three Goldstone bosons and a Higgs field. If we simply delete the
second term ($b=0$) then the symmetry breaking is not possible.

\subsection{The little group C$_{\mathbf{3}}$?}

The $\ell =3$ representation has three trivial singlet representations of $%
C_{3}$. In a particular choice of representations of $C_{3}$ generators the
fields $\chi _{1}$, $\chi _{2}$, and $\chi _{7}$ transform as a trivial
representation of $C_{3}$$^{.}$ When these fields take the non-zero
expectation values $\left\langle \chi _{a}\right\rangle =v_{a}\neq
0\;(a=1,2,7)$ while the others receive zero expectation values then (8a)
leads to the relations%
\begin{eqnarray}
v_{1}^{2}+v_{2}^{2} &=&-\frac{2a}{9(b-16c)}>0  \notag \\
v_{7}^{2} &=&-\frac{5a}{18(b-16c)}>0.
\end{eqnarray}%
the square-mass matrix (9) in this case is more complicated. Nevertheless
the eigenvalues of the matrix turn out to be exactly the same as those of
the matrix obtained in the case of tetrahedral group T, namely%
\begin{eqnarray*}
m_{1}^{2} &=&m_{2}^{2}=m_{3}^{2}=0,\quad m_{7}^{2}=-2a>0, \\
m_{4}^{2} &=&m_{5}^{2}=m_{6}^{2}=-\frac{40ac}{b-16c}>0\;
\end{eqnarray*}%
which imply, together with (33), the range of parameters $a<0,\quad
b-16c>0\quad and\;c>0.$ The minimum of the potential is also the same as
that of tetrahedral group%
\begin{equation}
V_{\min }=-\frac{a^{2}}{4(b-16c)}
\end{equation}

The $C_{3}$ being a subgroup of the tetrahedral group $T$ \ both groups have
the identical solutions. This follows from the fact that $\nu _{1}^{2}+\nu
_{2}^{2}+\nu _{7}^{2}=\nu _{4}^{2}=-\frac{a}{2(b-16c)}$. If one applies to
the solution for the little group $T$ a rotation leaving the above equation
invariant the result is precisely the same solution for $C_{3}$ . Therefore $%
C_{3}$ is not a little group of $SO(3)$ gauge symmetry for the potential of $%
\ell =3$ representation.

\subsection{The little group C$_{\mathbf{2}}$?}

In this case $\ell =3$ irreducible representation possesses three trivial
representations of $C_{2}$. In a special embedding of $C_{2}$ in $SO(3)$ we
can associate the fields $\chi $$_{3}$, $\chi _{4}$, and $\chi $$_{7}$ with
the trivial representations. By assigning the vacuum expectation values$%
\left\langle \chi _{a}\right\rangle =v_{a}\neq 0\;(a=3,4,7)$ we obtain the
solutions%
\begin{eqnarray}
v_{3}^{2}+v_{4}^{2} &=&-\frac{3a}{16(b+9c)}>0  \notag \\
v_{7}^{2} &=&-\frac{5a}{16(b+9c)}>0
\end{eqnarray}%
All gauge bosons gain masses. The square mass matrix (9) has the eigenvalues%
\begin{eqnarray}
m_{1}^{2} &=&m_{2}^{2}=m_{3}^{2}=0,\quad m_{4}^{2}=m_{5}^{2}=\frac{30ac}{%
(b+9c)}>0,  \notag \\
m_{6}^{2} &=&-\frac{45ac}{(b+9c)}>0,\;\;m_{7}^{2}=-2a>0
\end{eqnarray}%
(36) ad (37) are satisfied $a<0,\quad b+9c>0\;$and$\;c<0$ and the minimum of
the potential takes place at a value $V_{\min }=-\frac{a^{2}}{4(b+9c)}.$
Therefore $SO(3)\rightarrow C_{2}$ breaking produces the same results of $%
SO(3)\rightarrow D_{3}$ breaking except that the expectation values of the
fields are totally different. Absence of the quartic term ($c=0$) would
yield a theory with one Higgs fields and three Goldstone bosons. Also here,
we encounter with a problem similar to the case of $C_{3}-T$ symmetry. If
one applies a rotation to the solution for the little group $D_{3}$ one can
obtain the solution for the group $C_{2}$ where $\nu _{1}^{2}=\nu
_{3}^{2}+\nu _{4}^{2}+\nu _{7}^{2}=-\frac{a}{2(b+9c)}$. This proves that the 
$C_{2}$ is not a little group.

\section{Conclusions}

Breaking a gauged $SO(3)$ into its discrete subgroups has been studied with
renormalizable potentials of Higgs fields for the irreducible
representations $\ell =2$ and $\ell =3$. For the representation $\ell =2$
the little groups are $D_{\infty }$ and $D_{2}$. The third order potential
should be absent for a $D_{2}$ breaking while $D_{\infty }$ requires the
presence of all terms in the potential. Masses of Higgs and gauge bosons
have been calculated in terms of the parameters of the potential. An
interesting phenomenon is that in the breaking $SO(3)\rightarrow D_{2}$ one
finds a Goldstone boson as well as three Higgs particles, two of which have
equal masses.

Breaking $SO(3)$ into little groups of $\ell =3$ representation turns out to
be very interesting. For this representation third order potential term is
absent, instead one has two independent quartic polynomials in the
potential. One of the quartic term can be arranged as the square of the
quadratic term. We have calculated the Higgs particle masses as well as the
masses of the gauge bosons in terms of the parameters of the Lagrangian.
What we have observed is that the absence of the quartic polynomial which is
not square of the quadratic term leads to presence of Goldstone bosons.
Breaking a gauged $SO(3)$ into discrete subgroups without Goldstone bosons
are possible only with the presence of two independent quartic terms.\newline
What we have not dealt with is the nature of non-abelian cosmic strings
arising from such breakings. Since $SO(3)$ can be embedded in a GUT
candidate such as $E_{6}$ and $E_{8}$ through one of their $SU(3)$ subgroups
the technique we have discussed could be very useful for model builders, in
particular, for those who wish to study the role of non-abelian cosmic
strings in cosmological models. We had already checked that breaking $SU(3)$
through its $SO(3)$ subgroup with Higgs in its adjoint representation
results in abelian cosmic strings. This encourages us to study $E_{6}$
breaking from the point of view of cosmic string formation, a phenomena
which has not been worked out so far in the literature.

\section{Acknowledgement}

We are grateful to the referee for the substantial improvement of the paper.



\begin{thebibliography}{9}
\bibitem{1} Michel L,\textit{\ }1980\textit{\ Rev. Mod. Phys}., \textbf{52,}%
617.

\bibitem{2} de Gennes P E and Prost J, \textit{The Physics of Liquid Crystals%
}, \ 2$^{nd}$ edn. (oxford: Clarendon, 1993).\\[3pt]

\bibitem{3} Koca M, Ko\c{c} R and Al-Barwani M, 1997 \textit{J.Phys.A: Math.
Gen}., \textbf{30,} 2109.\\[4pt]

\bibitem{col} Coleman S, Aspects of Symmetry, \textit{Selected Erice lectures%
} (Cambridge: Cambridge University Press,1985).

\bibitem{4} Fel G L, 1995 \textit{Phys. Rev. E} \textbf{52,}702; Fel G L,
1995 \textit{Phys. Rev. E} \textbf{52,} 2692.\\[5pt]

\bibitem{5} Hindmarsh M B and Kibble T W B, 1995 \textit{Rep. Prog. Phys}., 
\textbf{58,} 477.\\[6pt]

\bibitem{6} O' Raifeartaigh L, Group Structure of Gauge Theories,
(Cambridge: Cambridge University Press, 1986).
\end{thebibliography}
\end{document}